\documentclass[final,5p,times,twocolumn]{elsarticle}

\usepackage{amssymb}
\journal{Chaos, Solitons and Fractals}

\begin{document}

\begin{frontmatter}

\title{Critical Slowing Down in a Real Physical System}

\author[inst1]{Mathias Marconi\corref{cor1}\fnref{label2}}

\affiliation[inst1]{organization={Université Côte d’Azur},
            addressline={CNRS}, 
            city={Institut de Physique de Nice},
            postcode={06200}, 
            state={Nice},
            country={France}}

\author[inst2]{Karin Alfaro-Bittner}
\author[inst1]{Lucas Sarrazin}
\author[inst1]{Massimo Giudici}
\ead{massimo.giudici@univ-cotedazur.fr}
\author[inst2]{Jorge Tredicce}

\affiliation[inst2]{organization={Universidad Rey Juan Carlos},
            addressline={Tulipán s/n}, 
            city={Móstoles},
            postcode={28933}, 
            state={Madrid},
            country={Spain}}

\begin{abstract}
The behavior of a dynamical system can exhibit abrupt changes when it crosses a tipping point. To prevent catastrophic events, it is useful to analyze indicators of the incoming bifurcation, as the divergence of the relaxation time of the system when approaching the critical point. However, this phenomenon, called critical slowing down (CSD), is hardly measurable in real physical systems. In this paper we provide experimental evidence of CSD in a laser system crossing the emission threshold and we analyze how it is affected by a time changing parameter and by noise.
\end{abstract}




\end{frontmatter}


\section{Introduction}
\label{sec:Intro}
Mankind has always been fascinated by the possibility of predicting the future. Precursor signs of the future were believed to appear in natural phenomena, as in the flight of birds (Ornithomancy), in the configuration of planets and stars (Astrology) and in the entrails of animals (Haruspicy) to cite few examples. Beyond magic and divination, common wisdom suggests to "Study the past if you would divine the future", to quote Confucius. In quantitative sciences, until the middle of the last century, the notion of predictability was associated to the knowledge of the equations ruling a dynamical system and of its initial condition. Unfortunately, this information is not available in real systems where noise is always present. Moreover, in 1963, Lorenz revealed the existence of chaotic systems, where determinism does not imply predictability \cite{Lorentz,Berge1987}. In the last fifty years, the question of what could be predicted in the evolution of real dynamical systems has attracted a great deal of attention in science. Great emphasis was placed on systems evolving towards a dramatic behavioral change \cite{Pomeau2012, Ott2014, Medeiros2017, Scheffer2018} in order to identify some precursor sign of the incoming bifurcation. The motivation of these studies is evident: such indicators would enable to avoid or, at least, to get prepared to catastrophic behavioral changes in real dynamical systems. 
In the eighties, the study of trajectories in phase space of a dynamical system during the transition between two stable solutions allowed to identify a phenomenon called critical slowing down (CSD) when points of marginal stability are approached \cite{Kramer1985}. Critical slowing down refers to the divergence of the relaxation time of a dynamical system when approaching a bifurcation and it is the key indicator for incoming behavioral change in all dynamical systems. Unfortunately, measuring CSD requires the possibility of perturbing a variable of the system and of analyzing its relaxation in time, which is often difficult to implement.
Hence, other indicators based on the time series of the dynamical system are often preferred, like the autocorrelation, the variance, the skewness, and the kurtosis \cite{Carpenter2006,Guttal2008,Scheffer2009,Dakos2012}. These indicators, called early warning signs (EWS) of an incoming bifurcation, tipping point or catastrophe, have been studied in various mathematical models representing a very wide variety of systems in different science area, including ecology \cite{Carpenter2006,Guttal2008}, climatology\cite{Ashwin2012}, physics \cite{Pomeau2012}, psychiatry \cite{Leemput2014,Wichers2016,Kunkels2023}, medicine \cite{Chen2019,Brien2021,Southall2021,Tredennick2022}, economy
\cite{Ismail2022} or neurology \cite{Meisel2015,Rings2019,Maturana2020}. Similar indicators have been also used in spatio-temporal systems \cite{Kwasniok2018,Donovan2022,Tirabassi2022,Veldhuis2022} or in networks \cite{Laren2023,George2023}.
Lately, other indicators were also sought, such as the evaluation of the spectrum of Lyapunov exponents which vary when approaching a tipping point \cite{Gallas2016, Nazarimehr2017, Bandy2021}. Other studies tried to obtain a normal form describing the incoming bifurcation through the statistical study of the variables~\cite{Ashwin2012, Pomeau2011, Boettiger2012}. Recently, a method based on machine learning was used to predict the existence of a drastic change in the behavior of a system \cite{Nonaka2023, Choi2022, Deb2022, Forzieri2022}.

However, these indicators are based on statistical analysis of the dynamic evolution of a given variable of the system whose variations may not necessarily linked to the proximity of a bifurcation. This lack of specificity in early warning signals have already been discussed in some papers \cite{Wilkat2019, Hastings2010, Lenton2011, Thompson2011, Remo2022, Boettner2022}.

\begin{figure*}[h!]
	\centering
	\includegraphics[scale=0.4,angle=0,origin=c]{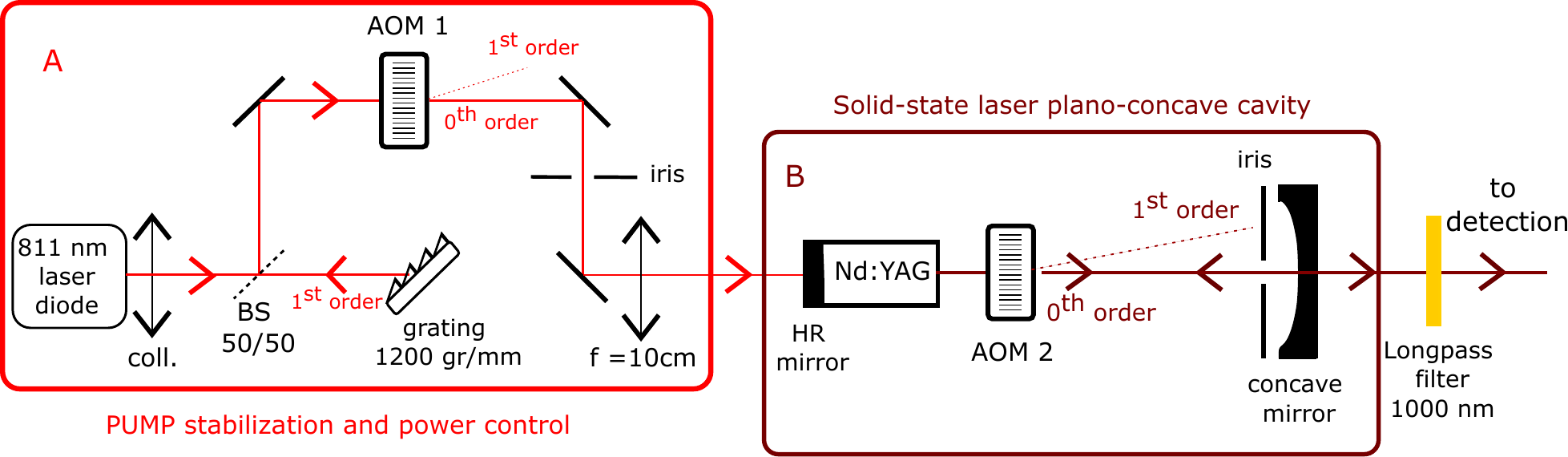}
	\caption{Laser platform for the measurement of CSD. The A section is the pump stabilization and output power control. AOM: Acousto-Optic Modulator, BS: Beam Splitter. Ramps of pump power with various speeds can be realised by an external control of the voltage applied to AOM1. The B section is the Nd:YAG laser cavity. Short square perturbations in the intracavity field are applied via the modulation of the voltage across AOM2. A longpass filter allows to eliminate the unwanted pumping light from the detection part. \label{fig:expsetup}}
\end{figure*}

 \begin{figure*}[h!]
 	\centering
 	\includegraphics[scale=0.27,angle=0,origin=c]{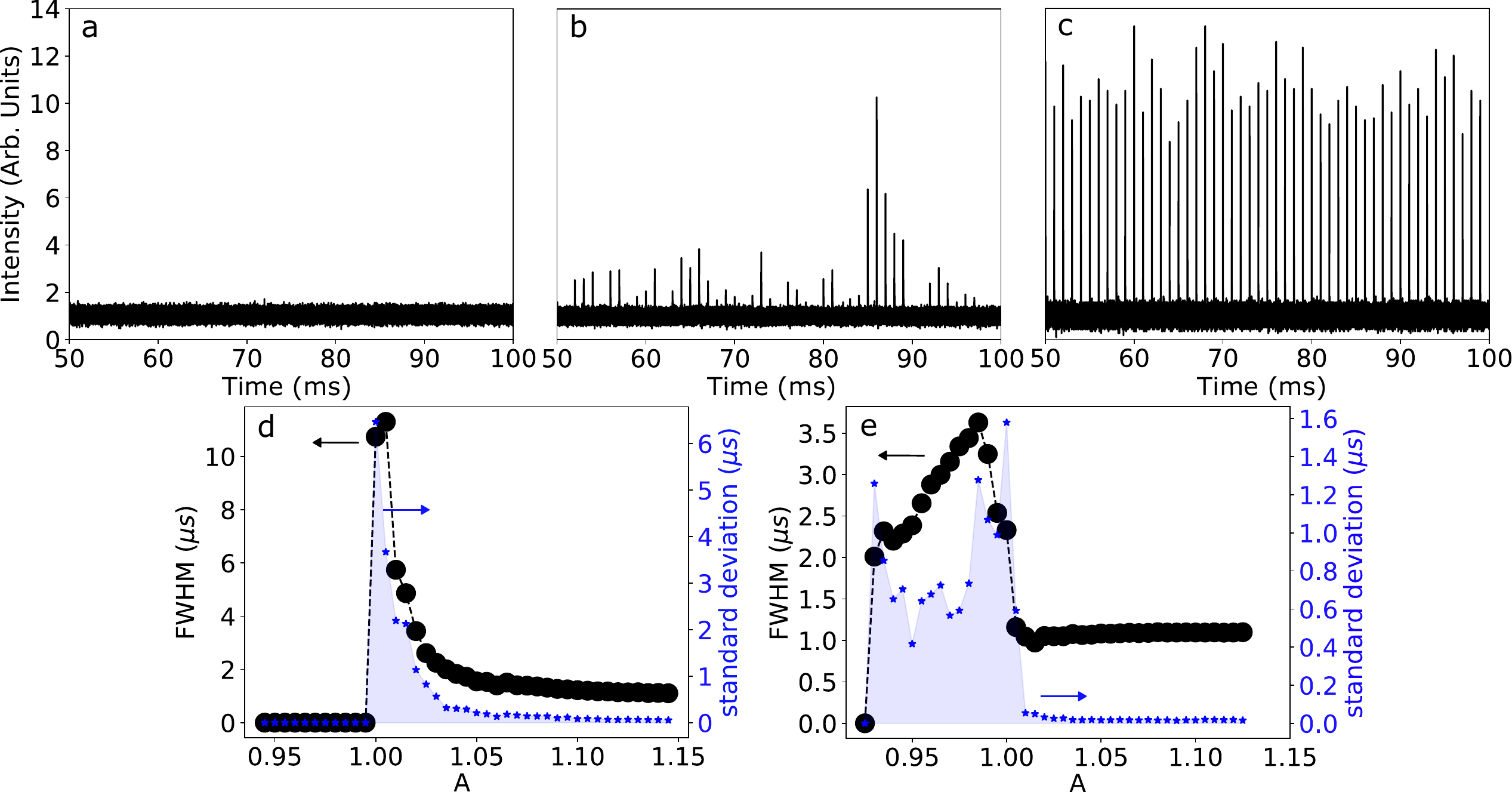}
 	\caption{a,b,c) Examples of time traces of the intensity pulses emitted by the laser (blue) when a sequence of 100 perturbations of 1 $\mu$s duration is applied to the cavity losses. The traces are obtained for a normalized pump power of respectively 0.99, 1.005 and 1.15. From these traces the average and maximum width of the pulse is estimated. d) Average (black trace) width of the laser response as a function of the normalized pump power $A$ and its variance (blue trace). The duration of the perturbation is 1 $\mu$s. e) Same as d) with a duration of the perturbation of 2.5 $\mu$s.
    \label{fig:exA_CSDstat}}
 \end{figure*}

A possible strategy for measuring directly CSD in real systems when the variables cannot be perturbed experimentally consists in perturbing an accessible control parameter. From this point of view, laser systems are a promising bench-test for CSD as several parameters are experimentally accessible and one of the variables, the intensity of the emitted electromagnetic field, can be readily measured by fast detectors and visualized on instruments. Class B lasers are dynamical systems with two active variables, the population inversion and the intensity output. They undergo to bifurcate when the pump parameter crosses the threshold and the nil solution of the intensity becomes unstable \cite{Tredicce2004, Pisarchik2002, Bonatto2017}. Previous attempts to measure CSD in a laser approaching threshold by perturbing the pump parameter failed \cite{Marconi2020} because this parameter is decoupled from the intensity variable below the bifurcation.

In this work we analyze experimentally and theoretically the response of a laser to a perturbation applied to a parameter which has a direct impact on both the variables of the system.  Accordingly, we give experimental evidence of CSD and we evaluate how it is affected by the presence of a time variable pump parameter that drives the laser towards the bifurcation. In addition, we show how the noise present in the system affects the measurement of CSD.

\section{Experimental evidence of the critical slowing down in a class-B laser}
\label{sec:experiment}
\subsection{Experimental setup}

 The experimental setup is displayed in Fig.~\ref{fig:expsetup}. We consider a laser composed by a Nd:YAG rode where the back plane facet HR coated at 1064 nm. A polished concave mirror having a 100 mm focal length is placed in front of the rode to define a plano-concave cavity having a length of 18 cm. The cavity contains an AOM crystal aligned in the cavity at the order zero. The AOM enables variation of the cavity losses from a bottom level given by the optical elements of the cavity and a maximum level when the AOM deviates the largest part of the beam toward the first order. The rode is pumped by a laser diode emitting at 811 nm whose intensity is controlled by a second AOM. The intracavity AOM is used to perturb the cavity losses with pulses having a duration of few microseconds and a repetition rate of 1 KHz. Timescales of the laser can be obtained from the relaxation oscillations characteristics and we obtain that population inversion recovery time is of the order of 200 $\mu$s, while the photon lifetime is about 200 ns.

\subsection{Measurement of the critical slowing for a stationary pump parameter}

In order to define an experimental scheme for revealing the occurrence of CSD, we analyze the response time of the laser intensity $\tau$ when perturbing the cavity losses. By definition, for a stationary situation of parameters, CSD occurs at the bifurcation point, i.e. at the laser threshold $A=A_{th}$. Then, the reliability of our experimental procedure can be tested accordingly. 
In Fig.\ref{fig:exA_CSDstat} we show a typical time trace of the intensity output of the laser when a set of periodic kicks having a duration of $\Delta t= 1 \mu$s are applied to the cavity losses for a fixed value of the pump parameter. 
While no response of the laser is observed far below threshold (panel a), a random response is observed close to threshold. Laser intensity pulses induced by a sudden variation of cavity losses are different in amplitude and in duration at every kick. These variations are maximal at $A=A_{th}$ (panel b)
and they tend to decrease when increasing the pump (panel c). The time series of Fig.~\ref{fig:exA_CSDstat} suggests that a statistical analysis of the laser intensity pulse is necessary for extracting a reliable indicator of the laser response.  In Fig.~\ref{fig:exA_CSDstat}d) we plot the average duration of the laser pulse intensity ($\tau$) as a function of the pump parameter together with its variance. Despite the spread of the laser response around threshold shown by the variance, the average value of $\langle \tau \rangle$, $\langle \tau \rangle$, exhibits a maximum value at $A=A_{th}$ and then it decreases as $A$ is increased above threshold. The variance $\tau$ decreases strongly above threshold and tends to negligeable values for $A>1.1 A_{th}$. This result suggests that the value of $A$ at which  $\langle \tau \rangle$ is maximum is a valuable indicator of CSD. The large value of the variance around the bifurcation point is quite common in dynamical systems with stochastic terms. Lasers feature several sources of noise both additive (spontaneous emission noise) and multiplicative (noise in the pump and/or in the cavity losses) which can affect its response to perturbations and smooth out the asymptotic growth of $\tau$ at the bifurcation point.  

An important ingredient of our experiment is the perturbation kick to the cavity losses. The measure of CSD requires a perturbation that can be considered as a sudden variation of the initial condition of the dynamical system without affecting its parameters. While this perturbation can be done simply in numerical simulation, in the experiment its implementation is critical. In our set-up, for technical limitations, the perturbation cannot be shorter than 0.2 $\mu$s. The results shown in Fig.~\ref{fig:exA_CSDstat}d) with $\Delta t =1 \mu$s  enables to locate CSD but they give no indications of the vicinity of the bifurcation point when the laser is in the off state. In order to extract a warning signal below the bifurcation point, we increase the kick duration to 2.5 $\mu$s and we show $\tau$ and its variance in fig. \ref{fig:exA_CSDstat}e). In this condition, the laser does respond to perturbations below threshold and $\langle \tau \rangle$ increases as $A$ approaches the threshold. $\langle \tau \rangle$ reaches a maximum slightly before the bifurcation point ($A=0.98 A_{th}$) and then it decreases abruptly to reach for a constant value at $A= A_{th}$ where also the variance becomes negligeable. Hence, for this kick duration, the maximum value of $\langle \tau \rangle$ fails in indicating the location of CSD. However, the increasing of $\langle \tau \rangle$ indicates the approaching of a bifurcation that will happen just after $\langle \tau \rangle$ reaches a maximum. 

To gain insights on the role of noise in the laser response and on the effects of kick duration, we compare our experimental findings with numerical simulations that include stochastic terms.
 
\section{Theoretical model and results}

\begin{figure}[h!]
	\centering
\includegraphics[scale=1,angle=0,origin=c]{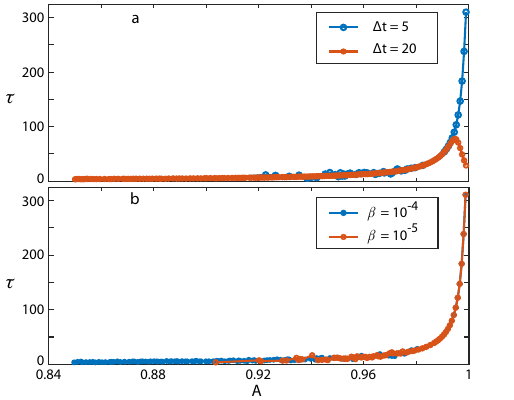}
	\caption{a) Average decay time $\langle \tau \rangle$ of the laser response as a function of the pump parameter $A$ for two different duration of the kicks and in presence of noise. $\Delta k=0.2$, $\beta =10^{-5}$.  b)  Average decay time $\langle \tau \rangle$ of the laser response as a function of the pump parameter $A$ for two noise levels. $\Delta k=0.2$, $\Delta t =5$. In both panels $\gamma = 10^{-4}$ and the average has been calculated using 100 events.\label{fig:NoRamA_theo}}
\end{figure}

A single mode Class B laser \cite{Arecchi1984} can be modeled by the following first-order differential equations:
\begin{eqnarray}
  \dot{E} =& -(1 - \Delta k+ \delta)E + E N + \beta, \\
  \dot{N} =& -\gamma [N - (A + \psi) + E^2 N],
\end{eqnarray}
where $E$ and $N$ are the electromagnetic field and the population inversion, respectively; $\gamma$ is the loss rate of $N$, and $A$ is the pump parameter. The time $t$ is normalized to the field loss rate. $\Delta k$ takes into account a perturbation in the loss rate of the field and $\beta$, $\delta$ and $\psi$ are random terms of zero average.
If $\Delta k$ and $\beta$ are neglected, the steady state solutions are $(E_1,N_1) = (0,A)$ and $(E_2,N_2) = (\sqrt{A - 1}, 1)$. A trivial linear stability analysis of the $(E_1,N_1)$ solution gives us the following eigenvalues $(\lambda_1,\lambda_2) = (A-1, -\gamma)$. $\lambda_1$ is associated to a perturbation on the field $E$ and $\lambda_2$ to a perturbation on the population inversion $N$. $(E_1,N_1)$ solution is stable for $A < 1$. At $A = 1$ a transcritical bifurcation takes place and, for $A > 1$ the solution $(E_2, N_2)$ becomes stable with eigenvalues $(\lambda'_1,\lambda'_2) = -\gamma A/2 \pm [\gamma^2 A^2 /4 - 2\gamma (A -1)]^{1/2}$.
Clearly, for both solutions, one eigenvalue vanishes at $A = 1$. Therefore there is a critical slowing down because at that point the decay time of a perturbation diverges. Following the method used in the experiment, we perturbed the cavity losses with short pulses. The choice of this parameter is explained by the equations describing the laser. If we consider the zero intensity solution, $(E_1, N_1)$, every change $\Delta k$ will induce a perturbation in the field values. The intensity ($I = E^2$) will then  decay to the steady state value and such decay will take longer as the parameter $A$ approaches the bifurcation point. It is worth remembering that a perturbation of pump parameter, as realized in \cite{Marconi2020}, does not affect the field value but only the variable $N$, which always decays in a time given by $\tau = 1/\gamma$. Then, CSD cannot be observed in this case. On the other hand, the CSD will always be observable as the bifurcation point is approached from the $(E_2,N_2)$ solution. 
As evidenced in the experimental observations, noise in the dynamical system can affect the laser response to a perturbation. We introduced into the model both multiplicative noise and additive noise in order to compare theoretical results with experimental ones. However, multiplicative noise had no relevant influence on the results, thus we focus on additive noise.

We realize numerically laser kicking by perturbing the losses through the parameter $\Delta k$. For a short time interval $\Delta t$, $\Delta k$ takes a finite value and we analyze the laser intensity pulse induced by this kick.  In presence of a stochastic term, the response of the laser is quite fluctuating and the degree of randomness depends on the noise level and on the distance of the parameter $A$ to the bifurcation value. Numerical time series are similar to the experimental ones and they suggest the need of a statistical analysis of the laser response.  In Fig.~\ref{fig:NoRamA_theo} we plot, as a function of the pump parameter $A$, the average value of the relaxation time of the laser intensity ($\langle \tau \rangle$) after a kick in the laser losses for different duration of the kicks $\Delta t$ and for different noise levels $\beta$. 
For a pulse duration of $\Delta t=5$, we observe that, regardless the presence of stochastic term, the relaxation time grows asymptotically in the proximity of the bifurcation point ($A=1$). This shows that noise level does not impact significantly the measure of CSD, provided that a statically analysis is performed on the system response. We repeat the above described numerical experiment with a kick duration of $\Delta t=20$. As in the experiment, when the perturbation duration exceeds a critical value, the maximum value of $\langle \tau \rangle$ occurs slightly before the bifurcation point and $\langle \tau \rangle$ decreases abruptly to reach a minimum at the bifurcation point. The analysis of the time series associated to this phenomenology enables to explain the effect of an exceedingly long perturbation pulse. 
The perturbation applied decreases the level of the cavity losses and therefore it decreases transiently the threshold value of the laser. If the perturbation is short enough, the laser does not integrate this change and the kick can be considered as a sudden variation of the initial condition rather than a change in the parameters. Hence, the laser intensity relaxes slowly and the maximum of the decay time happens at the position where the CSD occurs. If the perturbation is longer than a critical value, then the laser has the time to integrate the transiently threshold change and, when $A$ is close to $A_{th}$, it responds with a high intensity peak corresponding to the laser switch dynamics. Then, the (fast) decay time of the laser pulse is not related anymore with the relaxation process to the kick. Accordingly, the maximum of the decay time is reached before the point where the CSD must happen.

In conclusion, our experimental and numerical analysis of laser reveal that the noise present in a real system does not hinder the observation of CSD but it makes necessary multiple realizations of the system's perturbation and a statistical description of the response.

\subsection{Measurement of the critical slowing down with a time evolving parameter}
\begin{figure*}[h!]
	\centering
	\includegraphics[scale=0.4,angle=0,origin=c]{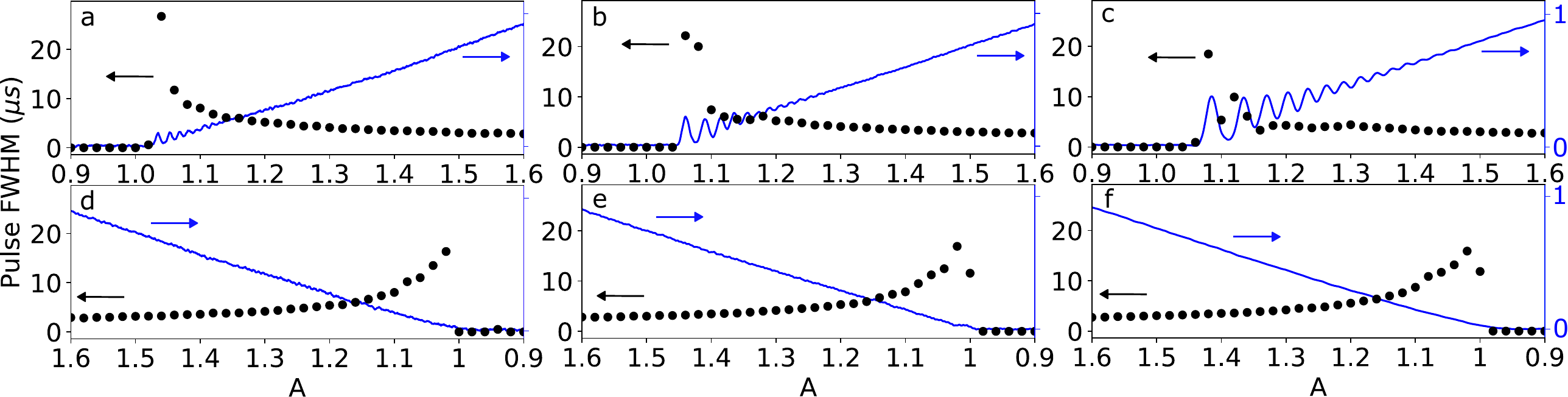}
	\caption{Laser response (blue curve) as the pump parameter is linearly increased with a speed $v$ (top row panels) or decreased with a speed $-v$ (low panel). Black spots: Average duration of the laser response to a kick of 1 $\mu s$ in the cavity losses when the perturbation is applied at different as a function of the values of the pump $A$ at which the perturbation is applied. a) $v=70 s^{-1}$ (50 Hz), b) $v=140 s^{-1}$ (100 Hz), c) $v=280 s^{-1}$ (200 Hz). d) $v=-70 s^{-1}$ (50 Hz), e) $v=-140 s^{-1}$ (100 Hz), f) $v=-280 s^{-1}$ (200 Hz). $\langle \tau \rangle$ has been measured by taking the average value of the pulse width over 10 events in panels a) and d), 20 events in panels b) and e) and 40 events in panels c) and f) \label{fig:rampe}}
\end{figure*}

The indications obtained above for measuring CSD in a real system when parameters are stationary can be used for measuring CSD in a system with time evolving parameters, a situation which is common in real world. In a recent paper ~\cite{Marconi2020} we have shown theoretically that, in these situations, CSD may take place for a pump parameter value, that we call $A_{CSD}$, beyond the threshold value $A_{th}$, thus making CSD an useless indicator for anticipating the incoming bifurcation. The shift between $A_{CSD}$ and $A_{th}$ depends on the variation speed of the evolving parameter \cite{Marconi2020} and it is shown analytically and numerically that CSD is taking place at the time value at which population inversion $N$ reaches the value of $N=1$. The offset between $A_{CSD}$ and $A_{th}$ increases with the speed of the ramp: $A_{c} \approx A_{th}(1+v/\gamma)$, where $v$ is the speed ramp in unit of $A/A_{th}$ normalized to the photon lifetime. This claim has not been experimentally demonstrated in \cite{Marconi2020} because the only accessible parameter for kicking the laser was its pump parameter and, as explained above, this is not useful for evidencing CSD. Here we exploit the results for static values of $A$ which indicate that CSD can be successfully evidenced by perturbing laser losses. 

To this aim, we modulate the pump parameter $A$ across the threshold value $A_{th}=1$ with a triangular function which increases and decreases linearly the pump ($A=A_0 \pm vt$). $A$ is varied in the interval $0.9 <A<1.6$ with variation speeds between $60s^{-1}$ and $500 s^{-1}$, which, normalized to photon lifetime, are speeds in the range $1.3\times 10^{-5}<v<1.1\times 10^{-4}$. During this modulation of $A$, the cavity losses are kicked by a short pulse applied at different values of the pump, i.e. at different positions on the ramp. This position can be controlled by changing the delay between the application of the perturbation and the beginning of the ramp. We evaluate the average duration of the laser response $\langle \tau \rangle$ for each kick position with respect to the pump value. As shown in Fig.~\ref{fig:expsetup}, $A$ parameter is modulated thanks to AOM which controls the power of the pump beam, while the intracavity AOM is used for kicking the cavity losses.
In Fig.\ref{fig:rampe} we plot $\langle \tau \rangle$ as the pump parameter $A$ is linearly evolving across the laser threshold $A_{th}=1$ for positive speeds [panels a), b) and c)] and for negative speeds [panels d), e) and f)].  The curve of the laser intensity output versus $A$ is also shown (blue curve). It is worth remembering that, when the pump is linearly increased with a speed $v$, the laser emission is delayed with respect to the time at which $A=A_{th}$, as shown experimentally   and theoretically in  \cite{Tredicce2004, Mandel1984, Scharpf1987}. Laser emission starts at a value of $A$, that we call "dynamical" threshold $A_{th,dyn}$,  whose offset with respect to $A_{th}$ increases with $v$, as we can observe in Fig.~\ref{fig:rampe}. Because of the energy accumulated, at $A=A_{th,dyn}$, the laser emits an intensity pulse followed by relaxation oscillations. Instead, for negative speed of pump change, i.e. when bifurcating from the lasing solution to the off solution, the offset between $A_{th}$ and the value of $A$ at which the laser switches off is negligible. This dynamical hysteresis has been already described and explained \cite{Tredicce2004}.
 Fig.~\ref{fig:rampe} shows that, for positive values of $v$, the maximum value of $\langle\tau\rangle$ is obtained when kicking the laser at the time for which the pump reaches the dynamical threshold, i.e., $A = A_{th,dyn}$. If the laser is kicked after $A = A_{th,dyn}$, $\langle \tau \rangle$ decreases progressively as the position of the perturbation in terms of $A$ increases with respect to $A = A_{th,dyn}$. For low speed of the ramp $v$, [(panel a)], there is no relevant difference between $A_{th}$ and $A_{th,dyn}$ but, as $v$ is increased (panels b) and c) $A_{th,dyn}$ takes place beyond the threshold value. Hence, the maximum value of $\langle \tau \rangle$, which indicates the position of CSD, occurs beyond the bifurcation point. The calculated values for $A_{CSD} $, according to \cite{Marconi2020}, are: $A_{CSD} \approx 1.016$ (panel a) $A_{CSD} \approx 1.032$ (panel b) $A_{CSD} \approx 1.064$ (panel c), in partial agreement with the values of $A_{CSD} $ displayed in  Fig.~\ref{fig:rampe}.
 For negative values of $v$, i.e. when the laser is crossing the bifurcation from the on state to the off state, there is no dynamical hysteresis and $A_{th} = A_{th,dyn}$. We observe that $\langle \tau \rangle$ progressively increases as the laser is kicked closer and closer to the time when laser switch off. It reaches a maximum for kicks at $A=1$ and it decreases for increasing delay between the kick and the time at which $A=1$. Hence, for negative speed of the ramp, CSD occurs at the $A=A_{th}$, as predicted theoretically and the increasing in the time $\langle \tau \rangle$ is a good indicator of the incoming bifurcation.

\begin{figure}[h!]
	\centering
	\includegraphics[scale=0.82,angle=0,origin=c]{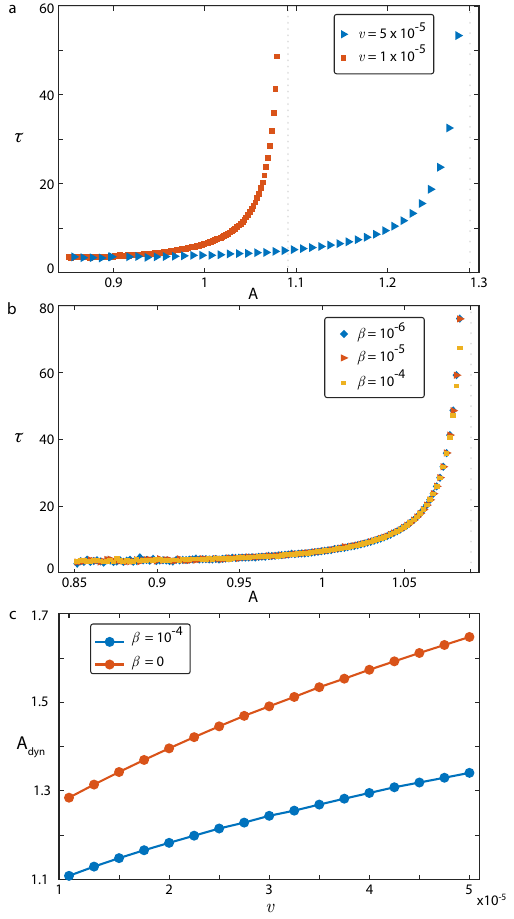}
	\caption{Average decay time $\langle \tau \rangle$ of the laser response to a kick in the cavity losses ($20\%$) when $A=A_0+vt$, $v>0$ as a function of $A$ for two values of $v$, $\beta=10^{-5}$ (panel a) and for different noise levels $\beta$ (panel b) for $v=10^{-5}$. Averaging has been performed on 100 events. Kick duration is $\Delta t=40$ and $\gamma = 10^{-4}$. Dashed lines indicate the value of $A$ at which CSD is supposed to occur by using the analytical formula. Panel c) position of $A_{th,dyn}$ as a function of the speed $v$ of change of $A$ in absence of stochastic term and for a finite noise level. \label{fig:Ramp-DiffAnoise}}
\end{figure}

In order to gain insights on the experimental observations, we implement numerical simulation using the model described above with a time evolving pump parameter: $A=A_0+vt$, for $v>0$.  In Fig.~\ref{fig:Ramp-DiffAnoise}, we plot the average relaxation time $\langle \tau \rangle$ of the laser response after a kick in the cavity losses in presence of the additive noise term $\beta$ and for different value of the ramp speed $v$. The value for $A$ for which CSD occurs ($A_{CSD}$) is calculated by using the analytical formula in \cite{Marconi2020} and Fig. \ref{fig:Ramp-DiffAnoise}a) shows that $\langle \tau \rangle$ indeed increases asymptotically when the kick is applied approaching $A=A_{CSD}$. As shown in Fig.~\ref{fig:Ramp-DiffAnoise}b), the presence of noise does not affect the value of $A$ at which $\langle \tau \rangle$ increased asymptotically, thus demonstrating the reliability of our measurement for the determination of $A_{CSD}$.

\begin{figure}[h!]
	\centering
	\includegraphics[scale=1,angle=0,origin=c]{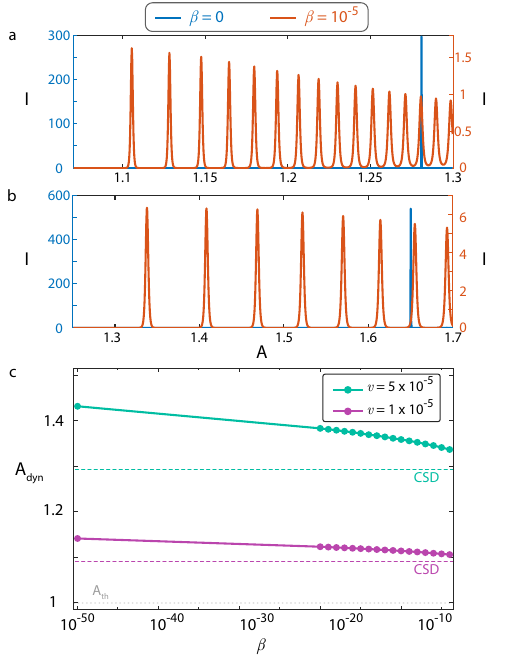}
	\caption{Laser intensity I as a function of pump A when the pump is linearly increased crossing the threshold value $A=1$ for different values of the speed $v$ at which A is changing: panel a) $v=1 \times 10^{-5}$, panel b) $v=5 \times 10^{-5}$. Panel c) $A_{th,dyn}$ calculated as a function of noise level for two speeds of the ramp (green curve $v=5\times 10^{-5}$, magenta curve $v=1\times 10^{-5}$). The position of $A_{CSD}$ is indicated by a dashed line.\label{fig:dynthreshold}}
\end{figure}

Noise affects instead the value of $A$ at which the laser starts to emit when $v>0$, i.e. the position of the dynamical threshold  $A_{th,dyn}$. This is shown in Fig. \ref{fig:Ramp-DiffAnoise}c and in Fig.~\ref{fig:dynthreshold} where we plot the intensity of the laser  as a function of the pump $A$ when the pump is linearly increased and it crosses the threshold value $A=1$. The time series in Fig.~\ref{fig:dynthreshold} compare with the experimental ones shown in Fig.~\ref{fig:rampe} and they show that the laser starts to emit with a large intensity peak followed by relaxation oscillations. The position of $A_{th,dyn}$ in absence of noise is indicated by a blue line, thus revealing how noise affects the value of $A_{th,dyn}$. In panel c) we analyze the position of $A_{th,dyn}$ as a function of the noise level for two different speeds. We notice how $A_{th,dyn}$ approaches $A_{CSD}$ as noise level is increased and this effect is more and more visible as the speed $v$ is increased.

\section{Conclusions}
 
Our analysis is devoted the possibility of predicting an incoming bifurcation in a dynamical system by measuring the decay time of a perturbation applied to one of the control parameters. Such decay time must diverge at a bifurcation point, according to the so-called critical slowing down phenomenon. However, in presence of an evolving parameter, CSD may occur beyond the bifurcation point, thus becoming unsuitable as early warning indicator. In this paper we have tested this circumstance in a low-dissipation dynamical system, namely a class B laser. We have shown that the noise present into the system makes a statistical analysis of the laser response to the perturbation compelling for measuring CSD. When the pump parameter of the laser is evolving in time across the laser threshold, the location of CSD with respect to the bifurcation point is different depending on the speed and on the direction of the parameter variation. While CSD occurs at the bifurcation point when the laser is switched off, it occurs at a pump value larger than the threshold when the laser is switched on and this gap increases with the speed of the pump variation. This result indicates that, in systems with a time evolving parameter, CSD may fail as early warning of an incoming bifurcation. Thanks to numerical simulations, we show that the value of the parameters at which the CSD happens is not affected by noise. However, noise affects significantly the pump value at which the dynamical threshold takes place. For typical noise levels in our laser experiment, this pump value tends towards the value at which CSD occurs.

 \bibliographystyle{elsarticle-num} 
 \bibliography{elsarticle}

\begin{thebibliography}{10}
\expandafter\ifx\csname url\endcsname\relax
  \def\url#1{\texttt{#1}}\fi
\expandafter\ifx\csname urlprefix\endcsname\relax\def\urlprefix{URL }\fi
\expandafter\ifx\csname href\endcsname\relax
  \def\href#1#2{#2} \def\path#1{#1}\fi

\bibitem{Lorentz}
E.~N. {Lorenz}, {Deterministic Nonperiodic Flow.}, Journal of the Atmospheric Sciences 20~(2) (1963) 130--148.

\bibitem{Berge1987}
P.~Berge, Y.~Pomeau, C.~Vidal, Order Within Chaos, Wiley, 1987.

\bibitem{Pomeau2012}
R.~D. Peters, M.~L. Berre, Y.~Pomeau, Prediction of catastrophes: An experimental model, Phys. Rev. E 86 (2012) 026207.

\bibitem{Ott2014}
T.~Nishikawa, E.~Ott, Controlling systems that drift through a tipping point, Chaos 24 (2014) 033107.

\bibitem{Medeiros2017}
E.~S. Medeiros, I.~L. Caldas, M.~S. Baptista, U.~Feudel, Trapping phenomenon attenuates the consequences of tipping points for limit cycles, Scientific Reports 7 (2017) 42351.

\bibitem{Scheffer2018}
S.~Bathiany, M.~Scheffer, E.~H. van Nes, M.~S. Williamson, T.~M. Lenton, Abrupt climate change in an oscillating world, Scientific Reports 8 (2018) 5040.

\bibitem{Kramer1985}
J.~Kramer, J.~Ross, Stabilization of unstable states, relaxation, and critical slowing down in a bistable system, The Journal of Chemical Physics 83~(12) (1985) 6234--6241.

\bibitem{Carpenter2006}
S.~Carpenter, W.~Brock, Rising variance: a leading indicator of ecological transition, Ecology Letter 9 (2006) 311--318.

\bibitem{Guttal2008}
V.~Guttal, C.~Jayaprakash, Changing skewness: an early warning signal of regime shifts in ecosystems, Ecology Letter 11 (2008) 450--460.

\bibitem{Scheffer2009}
M.~Scheffer, J.~Bascompte, W.~A. Brock, V.~Brovkin, S.~R. Carpenter, V.~Dakos, H.~Held, E.~H. van Nes, M.~Rietkerk, G.~Sugihara, Changing skewness: an early warning signal of regime shifts in ecosystems, Nature 461 (2009) 53--59.

\bibitem{Dakos2012}
V.~Dakos, S.~R. Carpenter, W.~A. Brock, A.~M. Ellison, V.~Guttal, A.~R. Ives, S.~Kéfi, V.~Livina, D.~A. Seekell, E.~H. van Nes, M.~Scheffer, Methods for detecting early warnings of critical transitions in time series illustrated using simulated ecological data, PLoS ONE 7~(7) (2012) e41010.

\bibitem{Ashwin2012}
P.~Ashwin, S.~Wieczorek, R.~Vitolo, P.~Cox, Tipping points in open systems: bifurcation, noise-induced and rate-dependent examples in the climate system, Phil. Trans. R. Soc. A 370 (2012) 1166--1184.

\bibitem{Leemput2014}
I.~A. van~de Leemput, M.~Wichers, A.~O.~J. Cramer, D.~Borsboom, F.~Tuerlinckx, P.~Kuppens, E.~H. van Nes, W.~Viechtbauer, E.~J. Giltay, S.~H. Aggen, C.~Derom, N.~Jacobs, K.~S. Kendler, H.~L.~J. van~der Maas, M.~C. Neale, F.~Peeters, E.Thiery, P.~Zachar, M.~Scheffer, Critical slowing down as early warning for the onset and termination of depression, PNAS 111 (2014) 87--92.

\bibitem{Wichers2016}
M.~Wichers, P.~C. Groot, Critical slowing down as a personalized early warning signal for depression, Psychotherapy and Psychosomatics 85 (2016) 114--116.

\bibitem{Kunkels2023}
Y.~K. Kunkels, A.~C. Smit, O.~Minaeva, E.~Snippe, S.~V. George, A.~M. van Roon, M.~Wichers, H.~Riese, Risk ahead: Actigraphy-based early-warning signals of increases in depressive symptoms during antidepressant discontinuation, Clinical Psychological Science 11~(5) (2023) 942--953.

\bibitem{Chen2019}
P.~Chen, E.~Chen, L.~Chen, X.~J. Zhou, R.~Liu, Detecting early‐warning signals of influenza outbreak based on dynamic network marker, J Cell Mol Med. 23 (2019) 395--404.

\bibitem{Brien2021}
D.~A. O’Brien, C.~F. Clements, Early warning signal reliability varies with covid-19 waves, Biol. Lett. 17 (2021) 20210487.

\bibitem{Southall2021}
E.~Southall, T.~S. Brett, M.~J. Tildesley, L.~Dyson, Early warning signals of infectious disease transitions: a review, J. R. Soc. Interface 18 (2021) 20210555.

\bibitem{Tredennick2022}
A.~Tredennick, E.~O’Dea, M.~Ferrari, A.~Park, R.~P, J.~Drake, Anticipating infectious disease re-emergence and elimination: a test of early warning signals using empirically based models., J. R. Soc. Interface 19 (2022) 20220123.

\bibitem{Ismail2022}
M.~Ismail, M.~Md~Noorani, M.~Ismail, F.~Abdul~Razak, Early warning signals of financial crises using persistent homology and critical slowing down: Evidence from different correlation tests, Front. Appl. Math. Stat. 8 (2022) 940133.

\bibitem{Meisel2015}
C.~Meisel, A.~Klaus, C.~Kuehn, D.~Plenz, Critical slowing down governs the transition to neuron spiking, PLoS Comput Biol 11 (2015) e1004097.

\bibitem{Rings2019}
T.~Rings, R.~von Wrede, K.~Lehnertz, Precursors of seizures due to specific spatial-temporal modifications of evolving large scale epileptic brain networks, Scientific Reports 9 (2019) 10623.

\bibitem{Maturana2020}
M.~I. Maturana, C.~Meisel, K.~Dell, P.~J. Karoly, W.~D’Souza, D.~B. Grayden, A.~N. Burkitt, P.~Jiruska, J.~Kudlacek, J.~Hlinka, M.~J. Cook, L.~Kuhlmann, D.~R. Freestone, Critical slowing down as a biomarker for seizure susceptibility, Nature Communications 11 (2020) 2172.

\bibitem{Kwasniok2018}
F.~Kwasniok, Detecting, anticipating, and predicting critical transitions in spatially extended systems, Chaos 28 (2018) 033614.

\bibitem{Donovan2022}
G.~Donovan, C.~Brand, Spatial early warning signals for tipping points using dynamic mode decomposition, Physica A 596 (2022) 127152.

\bibitem{Tirabassi2022}
G.~Tirabassi, C.~Masoller, Correlation lags give early warning signals of approaching bifurcations, Chaos, Solitons and Fractals 155 (2022) 111720.

\bibitem{Veldhuis2022}
M.~P. Veldhuis, R.~Martinez-Garcia, V.~Deblauwe, V.~Dakos, Remotely-sensed slowing down in spatially patterned dryland ecosystems, Ecography~(10) (2022) e06139.

\bibitem{Laren2023}
M.~NG, Kundu, M.~N., Early warnings for multi-stage transitions in dynamics on networks, J. R. Soc. Interface 20 (2023) 20220743.

\bibitem{George2023}
S.~V. George, S.~Kachhara, G.~Ambika, Early warning signals for critical transitions in complex systems, Physical Scripta 98 (2023) 072002.

\bibitem{Gallas2016}
M.~W. Beims, J.~A.~C. Gallas, Alignment of lyapunov vectors: A quantitative criterion to predict catastrophes?, Scientific Reports 6 (2016) 37102.

\bibitem{Nazarimehr2017}
F.~Nazarimehr, S.~Jafari, S.~M. R.~H. Golpayegani, J.~C. Sprott, Can lyapunov exponent predict critical transitions in biological systems?, Nonlinear Dynamics 88 (2017) 1493--1500.

\bibitem{Bandy2021}
D.~K. Bandy, E.~K.~T. Burton, J.~R. Hall, D.~M. Chapman, J.~T. Elrod, Predicting attractor characteristics using lyapunov exponents in a laser with injected signal, Chaos 31 (2021) 013120.

\bibitem{Pomeau2011}
Y.~Pomeau, M.~L. Berre, Critical speed-up vs critical slow-down: a new kind of relaxation oscillation with application to stick-slip phenomena, arXiv: 1107.3331v1 [physics.geo-ph] (2011).

\bibitem{Boettiger2012}
C.~Boettiger, A.Hastings, Quantifying limits to detection of early warning for critical transitions, J. R. Soc. Interface 9 (2012) 2527--2539.

\bibitem{Nonaka2023}
M.~Nonaka, M.~Agüero, M.~Kovalsky, Machine learning algorithms predict experimental output of chaotic lasers, Optics Letters 48 (2023) 1060--1063.

\bibitem{Choi2022}
J.~Choi, P.~Kim, Early warning for critical transitions using machine-based predictability, Mathematics 7 (2022) 20313–20327.

\bibitem{Deb2022}
S.~Deb, S.~Sidheekh, C.~Clements, N.~Krishnan, P.~Dutta, Machine learning methods trained on simple models can predict critical transitions in complex natural systems, R. Soc. Open Sci. 9 (2022) 211475.

\bibitem{Forzieri2022}
G.~Forzieri, V.~Dakos, N.~G. McDowell, A.~Ramdane, A.~Cescatti, Emerging signals of declining forest resilience under climate change, Nature 608 (2022) 534--556.

\bibitem{Wilkat2019}
T.~Wilkat, T.~Rings, K.~Lehnertz, No evidence for critical slowing down prior to human epileptic seizures, Chaos 29 (2019) 091104.

\bibitem{Hastings2010}
A.~Hastings, D.~B. Wysham, Regime shifts in ecological systems can occur with no warning, Ecology Letters 13 (2010) 464--472.

\bibitem{Lenton2011}
T.~M. Lenton, Early warning of climate tipping points, Nature Climate Change 1 (2011) 201--209.

\bibitem{Thompson2011}
J.~M.~T. Thompson, J.~Sieber, Predicting climate tipping as a noisy bifurcation: A review, International Journal of Bifurcation and Chaos 21 (2011) 399--423.

\bibitem{Remo2022}
F.~Remo, G.~Fuhrmann, T.~Jäger, On the effect of forcing on fold bifurcations and early-warning signals in population dynamics, Nonlinearity 35 (2022) 6485--6527.

\bibitem{Boettner2022}
C.~Boettner, N.~Boers, Critical slowing down in dynamical systems driven by nonstationary correlated noise, Physical Review Research 4 (2022) 013230.

\bibitem{Tredicce2004}
J.~R. Tredicce, G.~L. Lippi, P.~Mandel, B.~Charasse, A.~Chevalier, B.~Picque, Critical slowing down at a bifurcation, Am. J. Phys. 72 (2004) 799--809.

\bibitem{Pisarchik2002}
A.~N. Pisarchik, B.~F. Kuntsevich, Control of multistability in a directly modulated diode laser, IEEE Journ. of Quant. Electron. 38 (2002) 1594--1598.

\bibitem{Bonatto2017}
C.~Bonatto, A.~Endler, Extreme and superextreme events in a loss-modulated {CO}$_2$ laser: Nonlinear resonance route and precursors, Physical Review E 96 (2017) 012216.

\bibitem{Marconi2020}
M.~Marconi, C.~M{\'e}tayer, A.~Acquaviva, J.~M. Boyer, A.~Gomel, T.~Quiniou, C.~Masoller, M.~Giudici, J.~R. Tredicce, Testing critical slowing down as a bifurcation indicator in a low-dissipation dynamical system, Physical Review Letters 125~(13) (2020) 134102.

\bibitem{Arecchi1984}
F.~Arecchi, G.~Lippi, G.~Puccioni, J.~Tredicce, Deterministic chaos in laser with injected signal, Optics Communications 51~(5) (1984) 308--314.

\bibitem{Mandel1984}
P.~Mandel, T.~Erneux, Laser {L}orenz equations with a time-dependent parameter, Physical Review Letters 53~(19) (1984) 1818--1820.

\bibitem{Scharpf1987}
D.~B. C.~G. W.~Scharpf, M.~Squicciarini, J.~Tredicce, Experimental observation of a delayed bifurcation at the threshold of an {A}r+ laser, Optics Communications 63~(5) (1987) 344--348.

\end{thebibliography}

\end{document}